# Using a High-Throughput Screening Algorithm and relativistic Density Functional Theory to Find Chelating Agents for Separation of Radioactive Waste

Ashley Gannon,† Stephanie Marxsen,‡ Kevin Mueller,† Bryan D. Quaife,†,¶ and Jose L. Mendoza-Cortes*,‡,†,§,‖,⊥

†*Department of Scientific Computing, Florida State University, Tallahassee, Florida, 32306, USA*

‡*Department of Chemical & Biomedical Engineering, FAMU-FSU Joint College of Engineering, Tallahassee FL, 32310, USA*

¶*Geophysical Fluid Dynamics Institute, Florida State University, Tallahassee, Florida, 32306, USA*

§*Materials Science and Engineering, High Performance Materials Institute (HPMI), Florida State University, Tallahassee, Florida, 32310, USA.*

‖*Condensed Matter Theory, National High Magnetic Field Laboratory (NHMFL), Florida State University, Tallahassee, Florida, 32310, USA.*

⊥*Department of Physics, College of Arts and Science,Florida State University, Tallahassee FL, 32310, USA.*

E-mail: mendoza@eng.famu.fsu.edu

Phone: +(850) 410-6149. Fax: +(850) 410-6150

**Abstract**

Dangerous radioactive waste leftover from the Cold War era nuclear weapons production continues to contaminate sixteen sites around the United States. Although many challenges and obstacles exist in decontaminating these sites, two particularly difficult tasks associated with cleanup of this waste are extracting and separating actinide elements from the remainder of the solution, containing other actinide elements and non actinide elements. Developing effective methods for performing these separations is possible by designing new chelating agents that form stable complexes with actinide elements, and by investigating the interactions between the chelating agents and the actinide elements. In this work, new chelating agents (or ligands) with potential to facilitate the separation of radioactive waste are designed for Th, Pa, and U using relativistic Density Functional Theory (DFT) inconjunction with a high-throughput algorithm. We show that both methodologies can be combined efficiently to accelerate discovery and design of new ligands for separation of the radioactive actinides. The main hypothesis that we test with this approach is that the strength of secondary coordination sphere (SCS) can be tuned to increase the selectivity of binding with different actinides. More specifically, we show that links that connect two of the catecholamide ligands via covalent interactions are then added to increase the overall stability of the complex. The effect of increase in the selectivity is also observed when non-covalent interaction is used between ligands. We apply this approach for Th, Pa, and U, and discover linkers that can be used with other ligands. adding a butene link.

# Introduction

According to the report written by the US Department of Energy's Division of Basic Sciences in 2015, sixteen nuclear weapons production sites that operated during the Cold War still contain materials and wastes that are contaminated with radiation.[1] Three hundred million liters of highly radioactive waste, or enough waste to fill 120 Olympic sized swimming pools, are currently stored in tanks at just three of these sites. Numerous tanks have leaked over the past couple of decades, exposing the surrounding environment to some of the most dangerous and complicated materials that are known to exist on Earth.

One of the most onerous tasks associated with the cleanup of this nuclear waste is the extraction and separation of the radioactive actinide elements from the remaining non-radiation emitting elements. To facilitate the separation, new materials must be designed that are able to withstand the intense levels of radiation and acidity or alkalinity that are present in the nuclear waste. These new materials must also have a high selectivity for binding with radioactive actinide elements. Materials that have strong interactions with the metals are called chelating agents. Developing new chelating agents that possess a high selectivity for the actinides is only possible with a fundamental understanding of the interactions between the chelating agents and their intended binding elements. These interactions are known as host-guest interactions, where the ligands are the "hosts" and the metal center is the "guest".

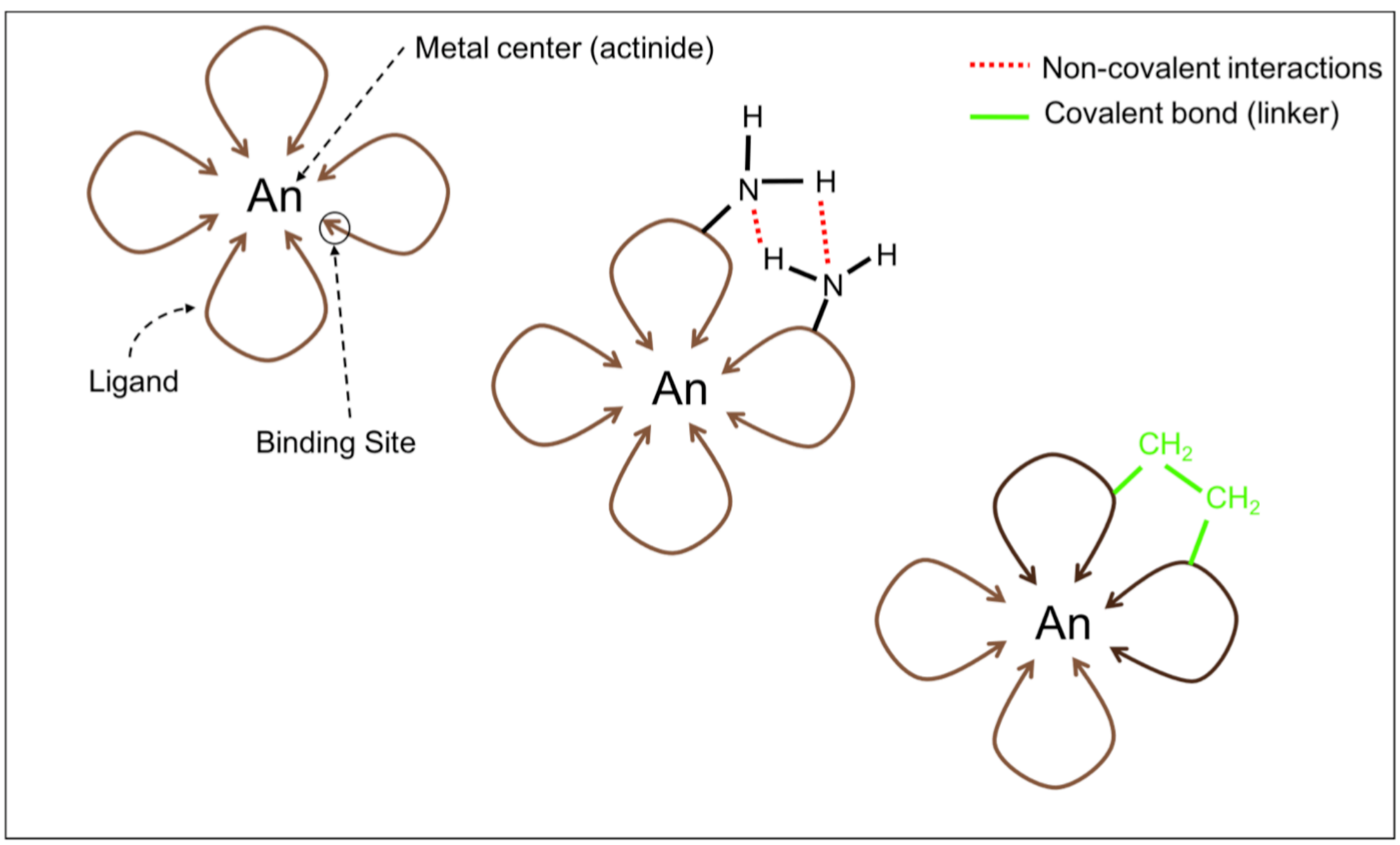

Figure 1: Ligand side interactions increase overall stability of the complex.

The nature of host-guest interaction has been the focus of research in many scientific fields. In coordination chemistry, host-guest complexes, or coordination compounds, are molecules with a metal center bound to one or more ligands. The study of interactions within coordination compounds has led to methods for ranking hosts in order of their affinity for binding with a certain guest. While these investigations are important for ligand design, they overlook an essential component: an efficient method for generating ligands.

Until recently, scientists have only been able to hypothesize binding site connections through graphical user interfaces or other forms of chemical intuition. Manually designing binding site connections can be time consuming, and the linking structures that form strong connections with the desired metal are usually not intuitive. In an attempt to reduce the time necessary for designing strongly-binding ligands, Hay *et al.* developed HostDesigner.[2] HostDesigner is a software package that identifies the optimal integrated ligands to attach to a desired metal using molecular mechanics algorithms. Once the algorithm creates these integrated complexes, the user must perform further calculations to investigate the stability of each complex and the nature of the interactions between the ligands and the metal produced.

HostDesigner is a high-throughput algorithm that can examine thousands of possible chemical structures in less than one minute, depending on the architecture of the complex. Once the user provides HostDesigner with an initial complex, the algorithm searches a library of 8,266 chelating agents for suitable links to connect the already existing ligands. The algorithm then ranks the chemical structures based on their relative energy. Complexes with the lowest relative energy must then be analyzed further with a more sophisticated level of theory, i.e. density functional theory with relativistic corrections, specifically, zeroth-order relativistic approximation (DFT-ZORA).

DFT-ZORA calculations describe the binding energy of the ligand-actinide structures and the coordinates and, in most cases, they are almost identical to experimental results. To determine if a given chelating agent is able to separate one actinide from other actinide elements, information on the relative binding energy of the chelating agent bound to different actinide elements is needed. Additionally, knowing the binding energy of many different chelating agents bound to a single actinide element allows ligands to be ranked based on their affinity for binding to that actinide.

Our main hypothesis is that ligands with secondary coordination sphere effects increase selectivity and thus separation, which can be used in the design of new ligands at accelerated pace using a high-throughout algorithm and first principles calculations. More specifically, in this work, new ligands are designed by investigating the second coordination sphere. As shown in Figure 1, the second coordination sphere exploits the use of non-covalent interactions and covalent bonds (linker formation) between ligands. The methodology involved in designing the new host-guest interactions includes the use of molecular mechanics and relativistic density functional theory (DFT-ZORA). We will explore the nature of the actinide complexes with non-interacting ligands, the non-covalent interactions between ligands, and the application of HostDesigner to design new linkers.

# Computational Methods

## Ligand Design Using HostDesigner

HostDesigner is a Fortran software package that uses molecular mechanics calculations that identify optimal integrated ligands for a specific metal. The algorithm predominantly used in this study is OVERLAY. The top half of Figure 2 shows an overview of the steps that HostDesigner's OVERLAY algorithm takes to design potential chelating agents for a user defined input structure that is chosen with, for example, Avogadro.[3] The bottom half of Figure 2 shows the more computationally expensive relativistic DFT as implemented in the Amsterdam Density Functional (ADF) that is applied to the top ranked HostDesigner output structures. We have modified slightly each algorithm for compatibility and convenience.

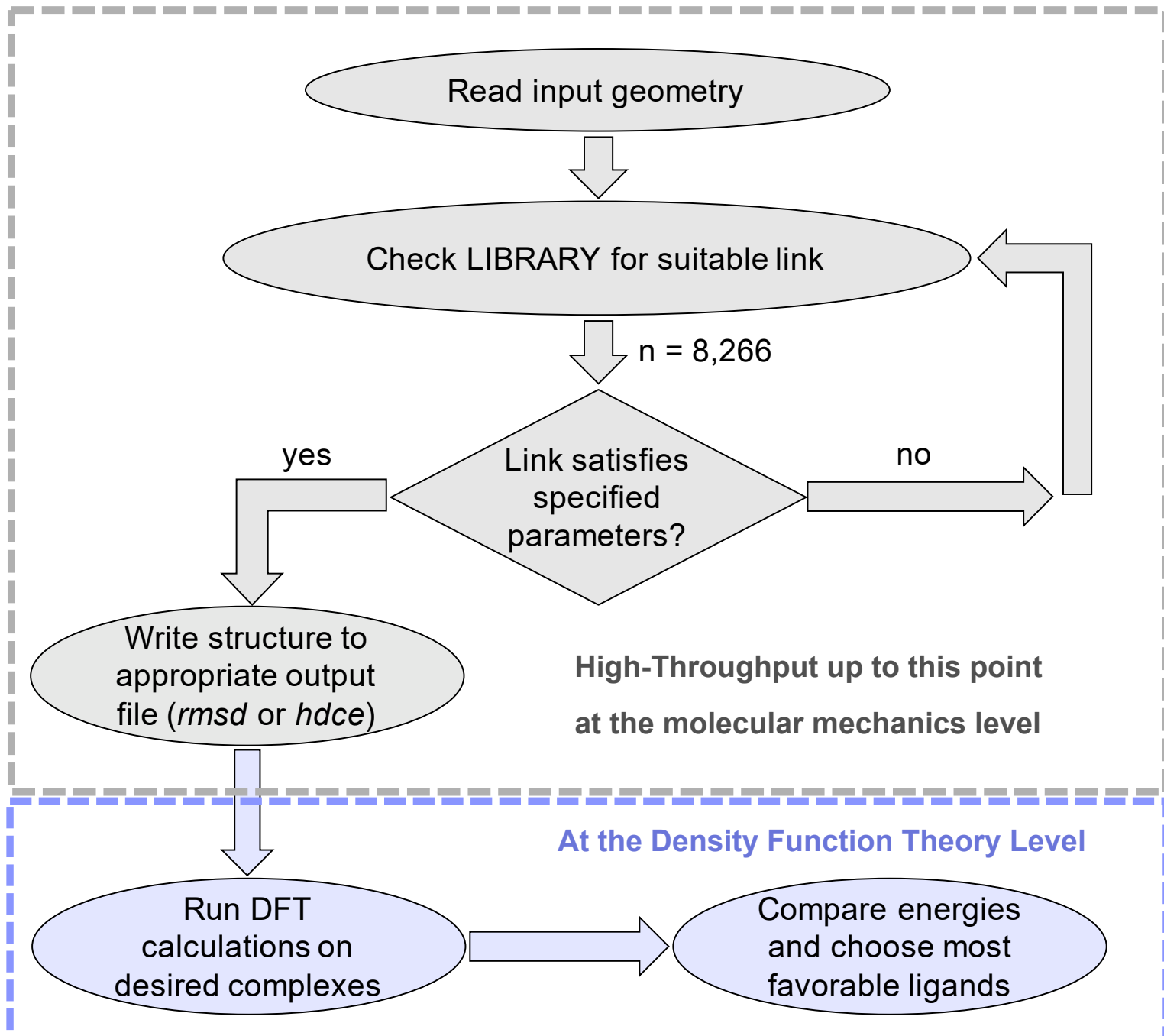

Figure 2: Basic algorithm used in this study: HostDesigner's OVERLAY is in upper portion while relativistic DFT is shown by the lower portion.

OVERLAY checks HostDesigner's library for a suitable link based on the input geometry. After examining the link and its connection to the host molecule, OVERLAY decides if the link satisfies the specified parameters. If there is no match, OVERLAY rejects the link. However, if there is a match, the coordinates of the link are added to those of the host molecule and saved for further analysis. Figure 3 shows a generic example of how OVERLAY connects two existing ligands to form a more stable complex. This process is repeated $n = 8,266$ times, for all links currently in the library.

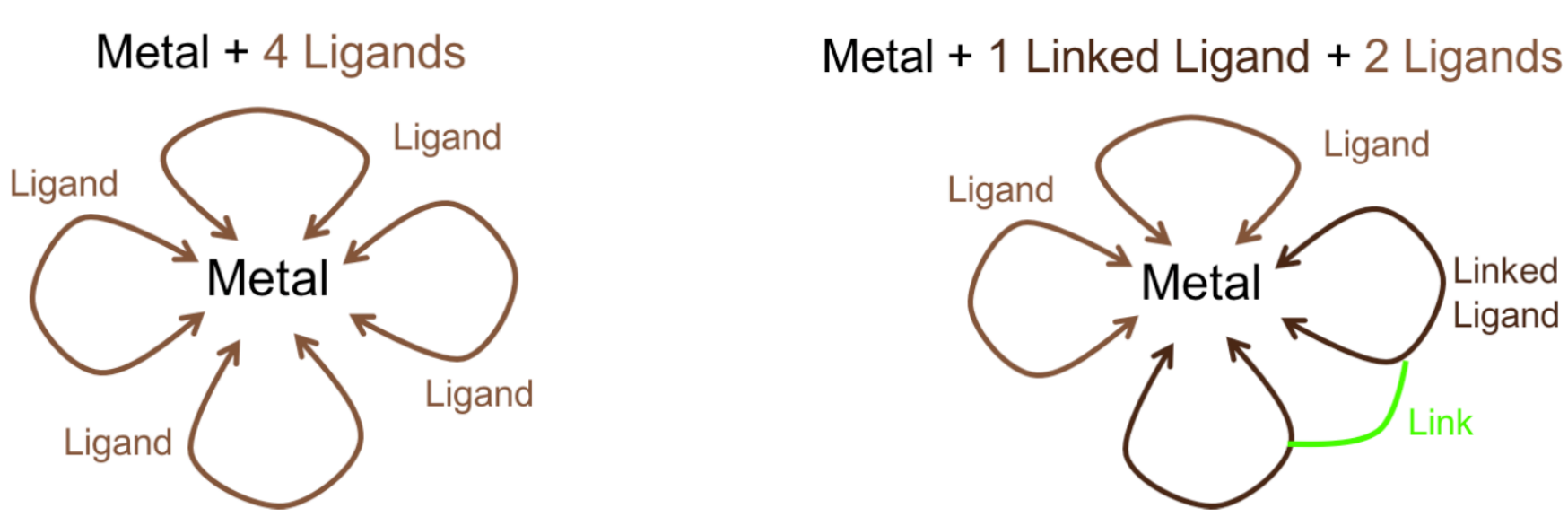

(a) Before HostDesigner (b) After HostDesigner

Figure 3: HostDesigner's OVERLAY algorithm adds the "link" shown in green in (b).

HostDesigner produces two sets of the (x,y,z) coordinates of the structures ranked according to root mean square deviation (rmsd) and HostDesigner conformational energy (hdce). A more in-depth explanation of ranking methods, and a step-by-step description of the OVERLAY algorithm can be found in the HostDesigner User's Manual.[4] HostDesigner is intended as a preliminary design tool for new host molecules. One limitation of HostDesiger is that it requires the user to be selective when choosing an input structure.

A more detailed analysis of output structures must follow the use of HostDesigner. Since the complexes generated by HostDesigner contain actinide elements, force field methods must be used with caution, as some parameters have not been developed with detail due to the lack of experimental data or the difficulty of using higher-order first principles methods on these atoms. Therefore, we use relativistic density functional theory (DFT) to carry out the refined analysis, combining accuracy and computational feasibility.

## Energy Calculations Using Relativistic DFT

We use density functional theory (DFT), as implemented in the Amsterdam Density Functional (ADF) version 2016-r51252.[5–7] All calculations are completed using the B3LYP-dUFF functional. B3LYP is a hybrid functional method, and dUFF accounts for dispersion corrections using the $C_6$ parameters from the universal force field (UFF). Relativistic triple zeta with polarization (ZORA:TZP) basis sets are used for all actinides, and double zeta with polarization (DZP) basis sets are used for the light elements (C, H, O, and N). Every basis set is all-electron, and the default numerical quality is used for integration and other convergence criteria. This level of theory will be referred to as B3LYP-dUFF/ZORA for the remainder of this work.

The binding energy of each complex is calculated with the optimized structures produced by B3LYP-dUFF/ZORA analysis. Equation (1) is used for this purpose:

$$\Delta E_{bind} = E_{Complex} - n(E_{Ligand}) - E_{LinkedLigand} - E_{Metal} \tag{1}$$

where,

$$\begin{aligned}
\text{Binding Energy}: &\quad \Delta E_{bind},\\
\text{DFT Energy of Complex}: &\quad E_{Complex},\\
\text{DFT Energy of Ligand}: &\quad E_{Ligand},\\
\text{DFT Energy of Linked Ligand}: &\quad E_{LinkedLigand},\\
\text{DFT Energy of Metal}: &\quad E_{Metal},\\
\text{Number of Ligands}: &\quad n.
\end{aligned}$$

An example of applying these equations to Figure 3 can be found in the supplementary information section.

Calculating the binding energy requires geometric optimization of the ligands, the linked ligand, the complex, and the actinide. Each of these values are computed in separate ADF calculations. However, the real interest is the difference between the binding energy of complexes with and without the linked ligand. Therefore, the binding energy of a complex designed by HostDesigner is reported as the difference between the binding energy of the complex with the linked ligand and the complex without the linked ligand. Using the generic complex observed in Figure 3, an example formula for the binding energy of the complex is:

$$\Delta E_{bind}^{diff} = \Delta E_{bind}^{linked}(b) - \Delta E_{bind}^{unlinked}(a) \tag{2}$$

Equation (2) can be generalized for all complexes generated by HostDesigner. For simplicity, $\Delta E_{bind}^{diff}$ will be referred to as $\Delta E$ for the remainder of this work.

## Applying HostDesigner to Recover Radioactive Waste

Determining the binding energy of various chelating agents allows us to determine the agents that provide the greatest degree of separation of an actinide element from a mixture of actinides and from non-radiation emitting elements. This is necessary when generating complexes for separating radioactive waste.

# Results and Discussion

The actinide-catecholamide complexes optimized with relativistic effects are used as input structures for HostDesigner. The geometry optimization calculations discussed in the methods section are performed so that the structures containing linked ligands obtained from HostDesigner's output files have a more accurate initial guess for the optimal coordinates. Even with an almost perfect guess of the first coordination sphere, optimizing the geometry of complexes containing more than 80 atoms is time consuming. Therefore, the first coordination sphere, the actinide element and eight oxygen atoms surrounding it, are "frozen" to simplify the model, meaning that the coordinates of the nine atoms are fixed while the remaining atoms in the structure are optimized normally. Using the resulting coordinates, the first coordination sphere is "unfrozen", and the full geometry optimization is performed.

## Pu(IV)-Nitrate and Pu(IV)-Water Complexes

Plutonium-nitrate, $[Pu(NO_3)_6]^{2-}$, and plutonium-water, $[Pu(H_2O)_9]^{4+}$, complexes are chosen as a basis to demonstrate that the binding energies and vibrational frequencies can be efficiently calculated for complexes containing actinide elements with results comparable to those found in literature using other computationally expensive methods. The original unoptimized structure of both complexes is obtained from the Cambridge Structural Database.[8] Figure 4 shows three-dimensional structures of both complexes.

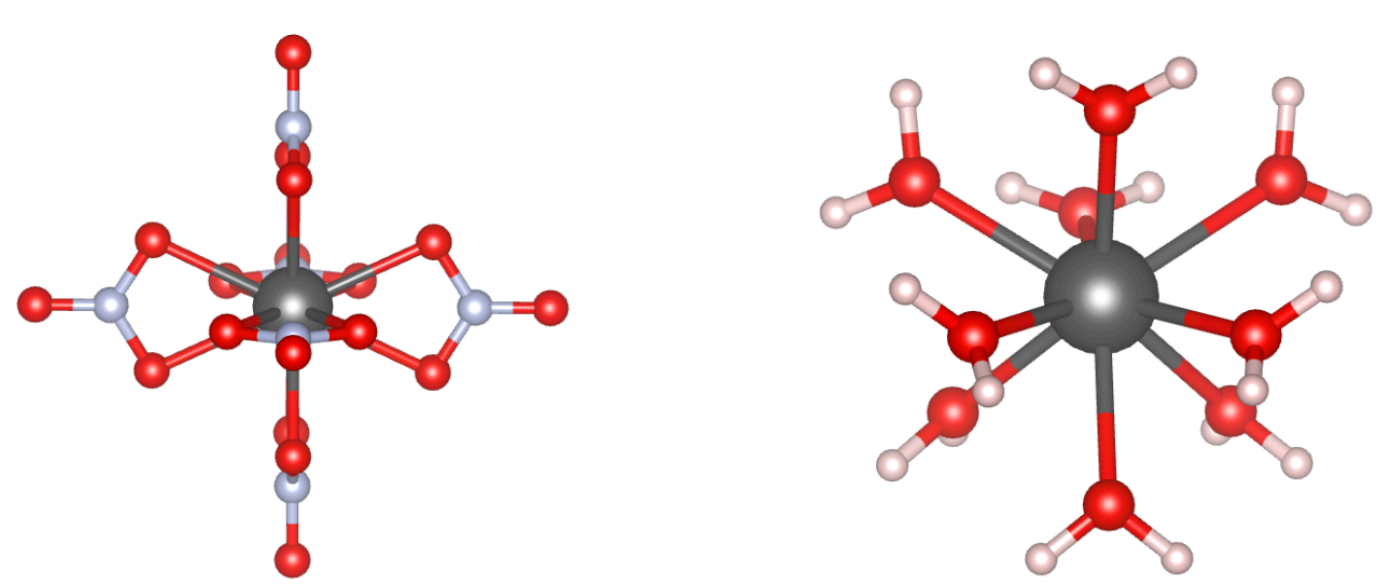

(a) $[Pu(NO_3)_6]^{2-}$ (b) $[Pu(H_2O)_9]^{4+}$

Figure 4: Three-dimensional structures of plutonium-nitrate and plutonium-water.

The geometries of both complexes are optimized in gas phase using B3LYP-dUFF/ZORA theory calculations. Table 1 lists three reactions, and the changes in binding energy (ΔE), changes in zero point energy (ΔZPE), changes in Gibbs free energy (ΔG), and the average plutonium-oxygen bond length for the optimized structures of these complexes. Note that the ΔZPE predicts the trend between ΔE and ΔG.

Table 1: Energy of plutonium-nitrate and plutonium-water complexes. The units are kcal/-mol for ΔE, ΔZPE and ΔG.

| Reaction | | | ΔE | ΔZPE | ΔG |
|---|---|---|---|---|---|
| $Pu(IV) + 6\,(NO_3)^-$ | $\longrightarrow$ | $[Pu(NO_3)_6]^{2-}$ | -1928.1 | 8.2 | -1858.1 |
| $Pu(IV) + 9\,(H_2O)$ | $\longrightarrow$ | $[Pu(H_2O)_9]^{4+}$ | -947.2 | 19.3 | -855.3 |
| $[Pu(H_2O)_9]^{4+} + 6\,(NO_3)^-$ | $\longrightarrow$ | $[Pu(NO_3)_6]^{2-} + 9\,(H_2O)$ | -980.9 | -11.1 | -1002.9 |

We validate the computational method by comparing our results to those produced by Wang *et al.*.[9] In their paper, experimental results obtained by Conradson *et al.* were listed in addition to their own computational results.[10,11] The calculations performed by Wang *et al.* were completed using the Gaussian 03 program, using B3LYP/6-311G(d,p) theory calculations for the light atoms, with added Relativistic Electron Core Potential (RECP) corrections for plutonium.

The calculations used by Wang are slightly different to the one used in this work. Table 2 shows the average Pu-O bond length determined in this work, by Wang *et al.*, and by Conradson *et al.*. Thus, the calculations completed in this work agree well with the other computational work, with only a 0.01Å difference (less than half a percent in relative error) in average bond length for each complex. Both computational methods are in good agreement with experimental results. However, the experimental results contains solvents and this might be the main reason for the small discrepancies since the theoretical results were calculated in the gas phase.

Table 2: Comparison of average Pu-O bond length to previous work.

| | This work (Å) | Wang *et al.* (Å) | Conradson *et al.* (Å) |
|---|---|---|---|
| $[Pu(NO_3)_6]^{2-}$ | 2.53 | 2.52 | 2.48 |
| $[Pu(H_2O)_9]^{4+}$ | 2.46 | 2.45 | 2.39 |

We calculate that the change in Gibbs free energy required to displace nine water molecules from plutonium and replace them with six nitrate molecules is ΔG = -1002.86 kcal/mol, whereas Wang *et al.* reported a value of ΔG = -968.5 kcal/mol, resulting in a difference of 35 kcal/mol, or an approximate relative error of 3.5%. Both methods reproduce the geometric structures well, however the present method, B3LYP-dUFF/ZORA, is expected to give more accurate energies and bond distances due to the consideration of explicit relativistic effects.

## Th(IV)

We start by optimizing the thorium-catecholamide complex using HostDesigner. Figures 5 and 6 contain the top twelve structures obtained from HostDesigner. There are several links that appear to be identical because of the two-dimensional illustrations, however, they have different configurations and this results in different binding energies. In particular, the four links rmsd3, rmsd4, hdce1, and hdce2; the two links rmsd6 and hdce4; and the two links rmsd5 and hdce3 all have different binding energies. This is because they correspond to different configurations, which the current approach also takes into account.

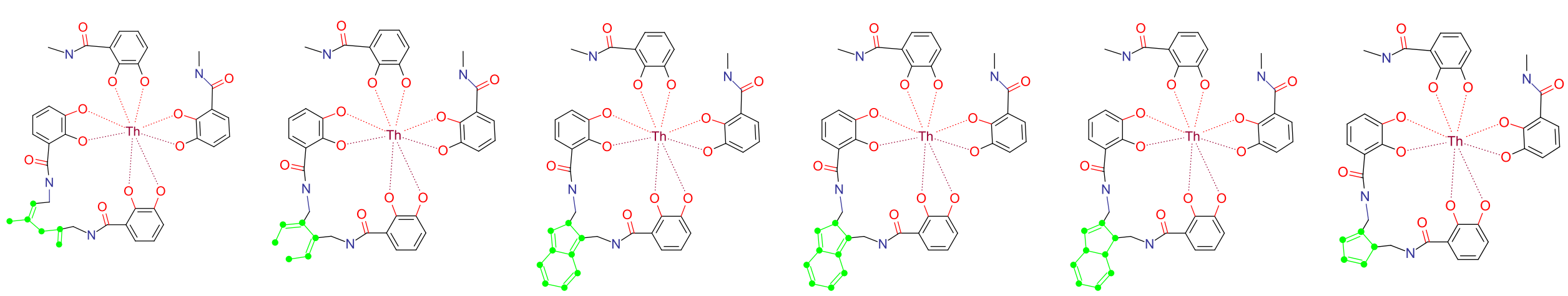

rmsd1 rmsd2 rmsd3 rmsd4 rmsd5 rmsd6

Figure 5: Top six thorium-catecholamide structures ranked by geometric parameters (rmsd). The link is highlighted in green on each structure.

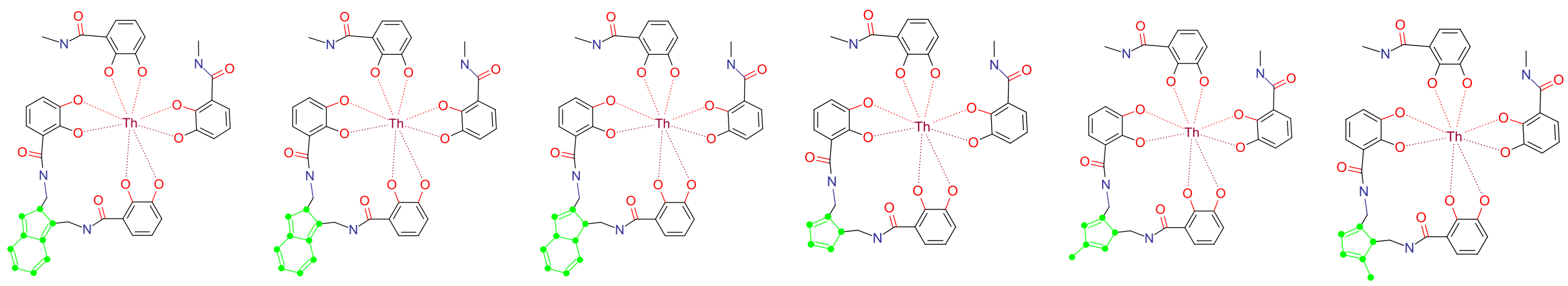

Figure 6: Top six thorium-catecholamide structures ranked by the conformational energy (hdce). The link is highlighted in green on each structure.

As described earlier, the geometry of each of the twelve complexes is further optimized using B3LYP-dUFF/ZORA calculations. Table 3 shows the name of all twelve links, the relative binding energy of the structure with the frozen first coordination sphere, the relative binding energy of the relaxed structure, and the average thorium-oxygen bond length of the relaxed structure. Additionally, the average thorium-oxygen bond length of the original, unlinked complex is shown. The binding energy of each complex is calculated based on the following reaction:

$$\mathrm{Th(IV)} + 2\,\mathrm{L}^{2-} + [\mathrm{L(x)}]^{4-} \longrightarrow [\mathrm{ThL_2L(x)}]^{4-}.$$

Here, L(x) represents rmsd 1-6 or hdce 1-6, and L is a single catecholamide ligand. Since the real interest is in the binding energy of each complex relative to the original unlinked complex, the reported binding energy values are relative to thorium-catecholamide. The relative binding energy is calculated using:

$$\Delta\mathrm{E} = \Delta\mathrm{E}([\mathrm{ThL_2L(x)}]^{4-}) \text{ - } \Delta\mathrm{E}([\mathrm{ThL_4}]^{4-}).$$

From this point on, the relative binding energy of the complex optimized with the first coordination sphere frozen will be called the frozen relative binding energy.

Table 3: Frozen and relaxed binding energy relative to unlinked thorium-catecholamide complex, and average actinide-oxygen bond length of the fully relaxed optimized geometry.

| Rank | Link | Frozen An-O $\Delta E$ (kcal/mol) | Relaxed An-O $\Delta E$ (kcal/mol) | Avg An-O (Å) |
|---|---|---|---|---|
| rmsd1 | 2-methyl-1,4-pentadiene | -101.52 | -101.72 | 2.463 |
| rmsd2 | cis,cis-2,4-hexadiene | -107.38 | -107.54 | 2.462 |
| rmsd3 | isoindene | -112.02 | -112.62 | 2.461 |
| rmsd4 | isoindene | -111.71 | -112.17 | 2.461 |
| rmsd5 | indene | -109.78 | -110.20 | 2.462 |
| rmsd6 | cyclopentadiene | -111.56 | -112.10 | 2.462 |
| hdce1 | isoindene | 111.90 | -112.38 | 2.461 |
| hdce2 | isoindene | -111.95 | -112.45 | 2.461 |
| hdce3 | indene | -109.78 | -110.20 | 2.462 |
| hdce4 | cyclopentadiene | -111.56 | -112.10 | 2.462 |
| hdce5 | 2-methylcyclopentadiene | -111.84 | -112.29 | 2.462 |
| hdce6 | 1-methylcyclopentadiene | -110.11 | -110.70 | 2.462 |
| | Unlinked Ligands | — | — | 2.463 |

For all complexes, the relaxed relative binding energy is within 0.6 kcal/mol of the frozen relative binding energy. Additionally, the relative binding energy of every complex decreased when the geometry is relaxed. The average thorium-oxygen bond length found in the relaxed complexes is in general slightly shorter than the original unoptimized complex, at most by 0.002 Å.

The effect that each link has on the stability of the original complex is more apparent when the relaxed relative binding energy is shown graphically. Figure 7 shows two graphs of the binding energy of each of the structures from Table 3. The rmsd complexes are presented separately from the hdce complexes for clarity. From these graphs, the most stable complex is created with the addition of rmsd3, isoindene. The rmsd4, hdce1, and hdce2 links, also isoindene, have relative binding energies within 0.5 kcal/mol of the rmsd3 isoindene link. Therefore, while HostDesigner may have placed the links in different orientations, all four complexes optimize to the same configuration at the B3LYP-dUFF/ZORA level of theory. A similar phenomenon occurs with both rmsd5 and hdce3 (indene) and rmsd6 and hdce4 (cyclopentadiene).

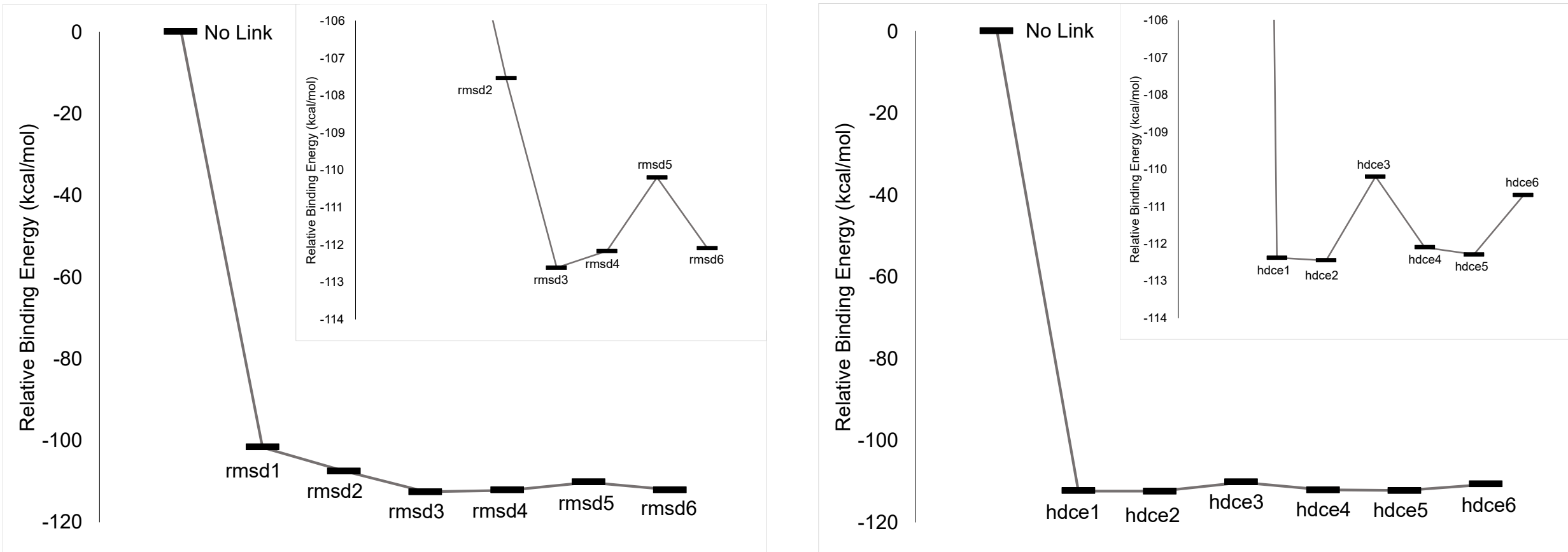

Figure 7: Relaxed binding energy of the top twelve thorium complexes produced by HostDesigner relative to the original, unlinked thorium-catecholamide complex.

## Pa(IV)

Protactinium-catecholamide is the next optimized actinide-catecholamide complex used as the input to HostDesigner. Figures 8 and 9 contain the top twelve structures generated by HostDesigner. The geometries of all twelve structures are optimized using B3LYP-dUFF/ZORA theory calculations. The binding energy of each complex is calculated based on:

$$\mathrm{Pa(IV)} + 2\,\mathrm{L}^{2-} + [\mathrm{L(x)}]^{4-} \longrightarrow [\mathrm{PaL_2L(x)}]^{4-}.$$

The relative binding energy is calculated using:

$$\Delta E = \Delta E([\mathrm{PaL_2L(x)}]^{4-}) - \Delta E([\mathrm{PaL_4}]^{4-}).$$

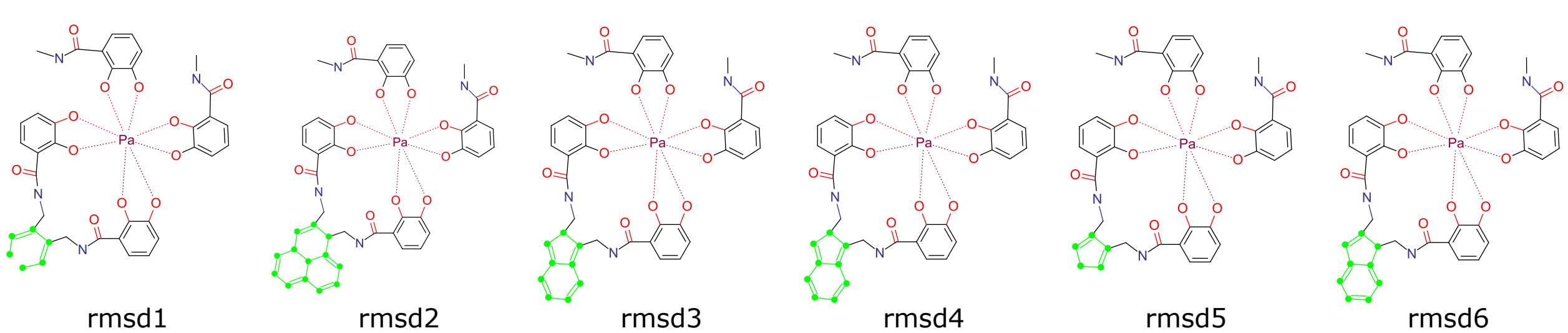

Figure 8: Top six protactinium-catecholamide structures ranked by geometric parameters (rmsd). The link is highlighted in green on each structure.

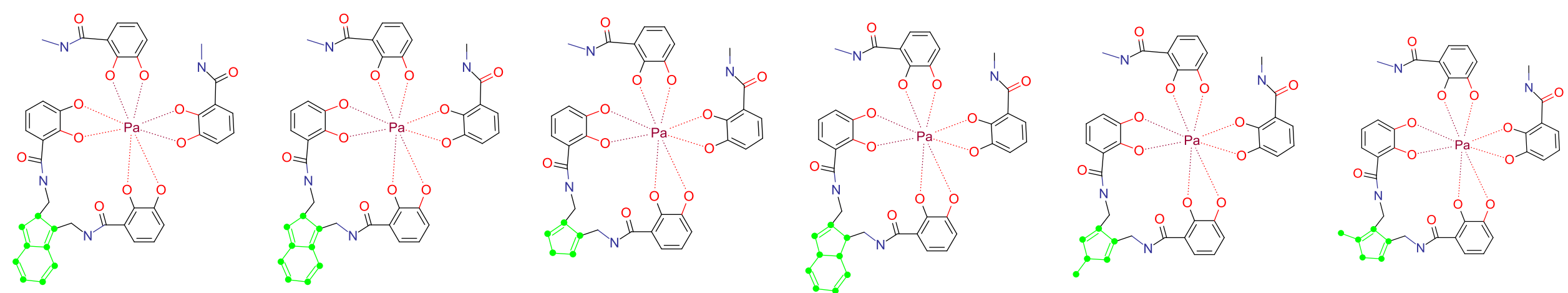

hdce1 hdce2 hdce3 hdce4 hdce5 hdce6

Figure 9: Top six protactinium-catecholamide structures ranked by estimated conformational energy (hdce). The link is highlighted in green on each structure.

Table 4 shows the name of all twelve links, the relative binding energy of the structure with the frozen first coordination sphere, the relative binding energy of the relaxed structure, and the average protactinium-oxygen bond length of the relaxed structure. Additionally, the average protactinium-oxygen bond length of the original, unlinked complex is reported.

Table 4: Frozen and relaxed binding energy relative to unlinked protactinium-catecholamide complex, and average actinide-oxygen bond length of the fully relaxed optimized geometry.

| Rank | Link | Frozen An-O $\Delta E$ (kcal/mol) | Relaxed An-O $\Delta E$ (kcal/mol) | Avg An-O (Å) |
|---|---|---|---|---|
| rmsd1 | cis,cis-2,4-hexadiene | -78.22 | -81.38 | 2.431 |
| rmsd2 | phenalene | -81.90 | -94.91 | 2.332 |
| rmsd3 | isoindene | -98.20 | -110.80 | 2.332 |
| rmsd4 | isoindene | -98.20 | -110.82 | 2.332 |
| rmsd5 | cyclopentadiene | -82.90 | -83.96 | 2.433 |
| rmsd6 | indene | -83.64 | -87.47 | 2.332 |
| hdce1 | isoindene | -113.33 | –125.94 | 2.332 |
| hdce2 | isoindene | -97.79 | -110.41 | 2.332 |
| hdce3 | cyclopentadiene | -104.89 | -106.07 | 2.434 |
| hdce4 | indene | -101.66 | -105.71 | 2.332 |
| hdce5 | 3-methylcyclopentadiene | -82.86 | -84.00 | 2.433 |
| hdce6 | 2-methylcyclopentadiene | -108.35 | -109.34 | 2.433 |
| | Unlinked Ligands | — | — | 2.434 |

Unlike the thorium complexes, the relative binding energy of the frozen and relaxed complexes containing protactinium differed significantly in some cases. For example, all of the complexes with the isoindene link (rmsd3, rmsd4, hdce1, hdce2) stabilize around 12.6 kcal/mol, when the first coordination sphere is allowed to relax. The greatest difference is in the rmsd2 link, phenalene, at 13.01 kcal/mol. Additionally, the average protactinium-oxygen bond length of many complexes contract by more than 0.1 Å.

Figure 10 is a graphical representation of the relaxed relative binding energy of the complexes shown in Table 4. The most stable complex is created with the addition of hdce1, isoindene. Note that in this case, unlike with thorium, the other isoindene complexes (rmsd3, rmsd4, and hdce2) do not have the same relative binding energy as hdce1. The isoindene complexes from rmsd3, rmsd4, and hdce2 all have the same binding energy, but that value is approximately 15 kcal/mol more than the one from hdce1. This is likely due to a different configuration of the isoindene link. Additionally, neither the indene complexes (rmsd6 and hdce4) nor the cyclopentadiene complexes (rmsd5 and hdce3) optimize to configurations with the same relative binding energy. The indene complexes differ by more than 18 kcal/mol, while the cyclopentadiene complexes differ by more than 22 kcal/mol.

Finally, it is worth mentioning that the range of relative binding energy is much larger for protactinium complexes than it is for thorium complexes. The difference in binding energy between the most and least stable thorium complexes is less than 11 kcal/mol, while for the protactinium complexes, the same difference is nearly 45 kcal/mol.

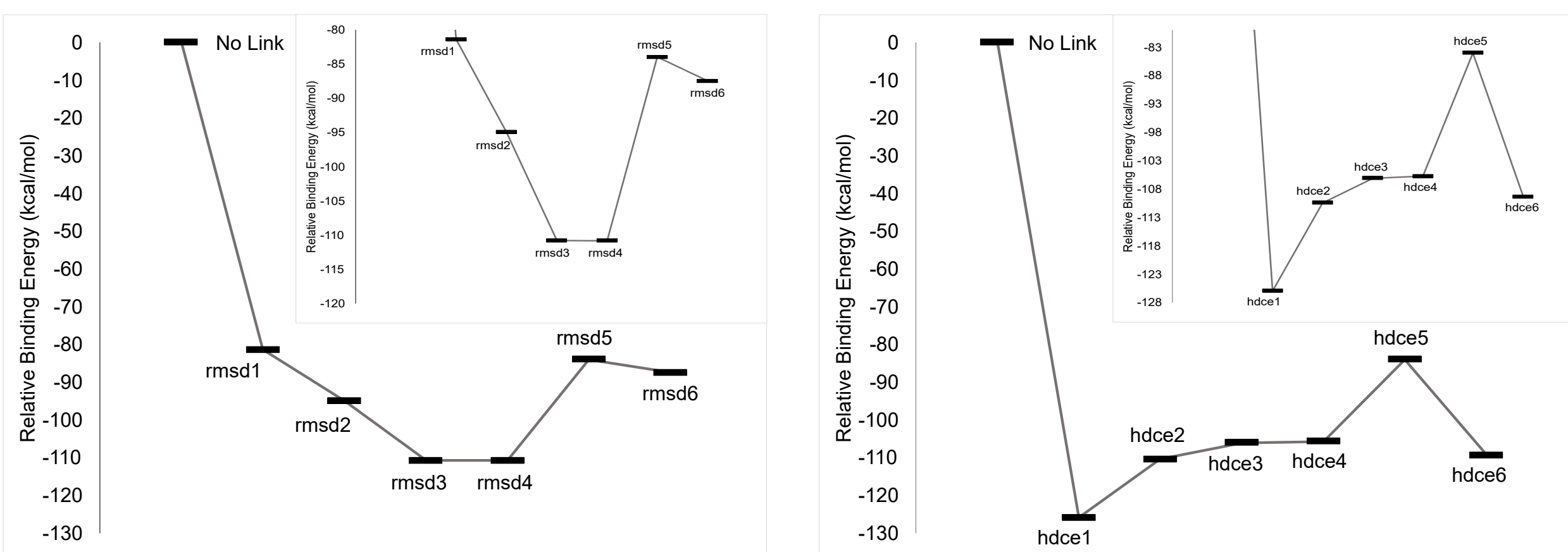

Figure 10: Relaxed binding energy of the top twelve protactinium complexes produced by HostDesigner relative to the original, unlinked protactinium-catecholamide complex.

## U(IV)

The optimized uranium-catecholamide complex is used as the input to HostDesigner. Figures 11 and 12 contain the top twelve structures from HostDesigner. The geometry of each structure is optimized using calculations at the B3LYP-dUFF/ZORA level of theory. The binding energy of each complex are calculated based on the reaction:

$$\mathrm{U(IV)} + 2\,\mathrm{L}^{2-} + [\mathrm{L(x)}]^{4-} \longrightarrow [\mathrm{UL_2L(x)}]^{4-}.$$

The relative binding energy is calculated using:

$$\Delta E = \Delta E([\mathrm{UL_2L(x)}]^{4-}) - \Delta E([\mathrm{UL_4}]^{4-}).$$

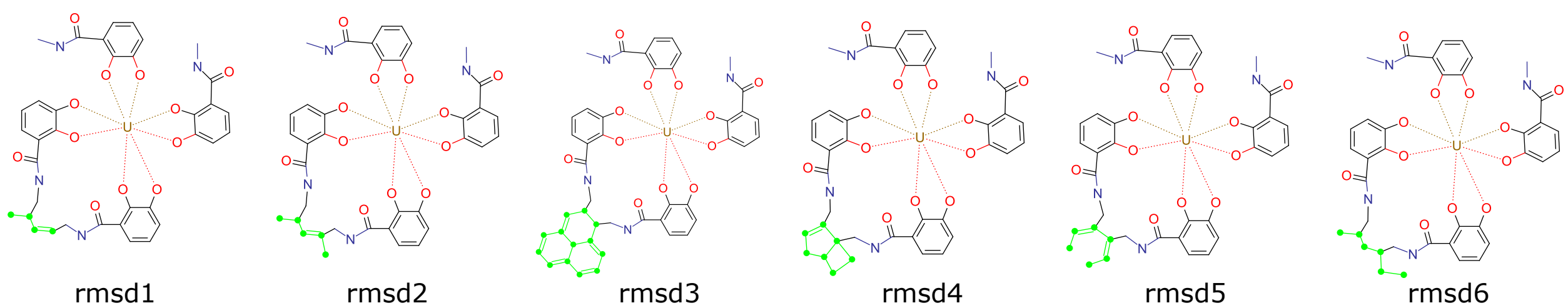

Figure 11: Top six uranium-catecholamide structures ranked by geometric parameters (rmsd). The link is highlighted in green on each structure.

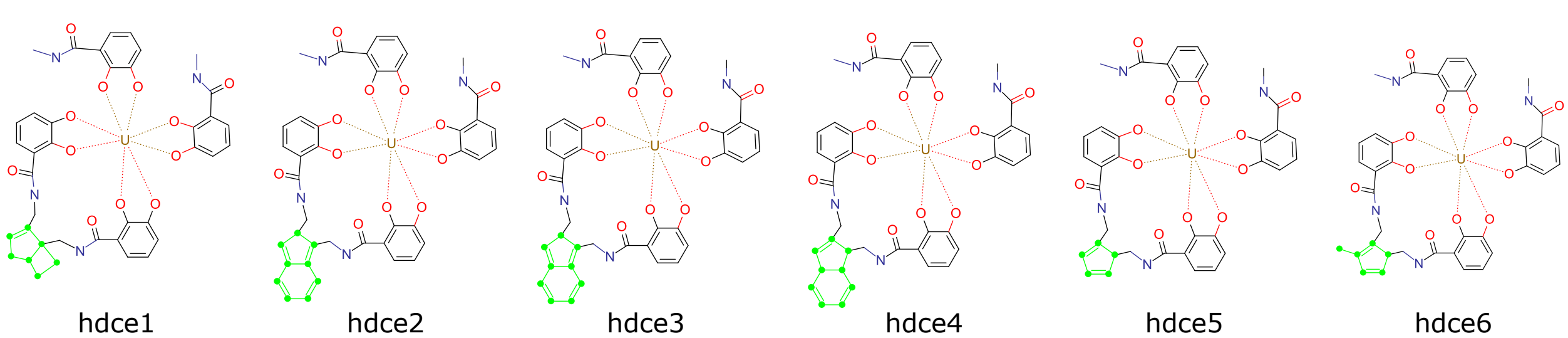

Figure 12: Top six uranium-catecholamide structures ranked by estimated conformational energy (hdce). The link is highlighted in green on each structure.

Table 5 shows the name of each link, the relative binding energy of the structure with its first coordination sphere frozen, the relative binding energy of the relaxed structure, and the average uranium-oxygen bond length of the relaxed structure. The average uranium-oxygen bond length of the original, unlinked complex is reported.

Table 5: Frozen and relaxed binding energy relative to unlinked uranium-catecholamide complex, and average actinide-oxygen bond length of the fully relaxed optimized geometry.

| Rank | Link | Frozen An-O $\Delta E$ (kcal/mol) | Relaxed An-O $\Delta E$ (kcal/mol) | Avg An-O (Å) |
|---|---|---|---|---|
| rmsd1 | butene | -100.82 | -101.53 | 2.417 |
| rmsd2 | trans-2-pentene | -61.79 | -63.53 | 2.414 |
| rmsd3 | phenalene | -63.69 | -60.60 | 2.417 |
| rmsd4 | bicyclo[2.2.1]heptene | -65.09 | -65.49 | 2.413 |
| rmsd5 | cis,cis-2,4-hexadiene | -61.34 | -61.72 | 2.411 |
| rmsd6 | trans-1,3-hexadiene | -55.13 | -61.08 | 2.414 |
| hdce1 | bicyclo[2.2.1]heptene | -65.10 | -65.50 | 2.413 |
| hdce2 | isoindene | -46.43 | -47.43 | 2.369 |
| hdce3 | isoindene | -46.11 | -47.21 | 2.368 |
| hdce4 | indene | -44.79 | -42.15 | 2.415 |
| hdce5 | cyclopentadiene | -40.86 | -42.62 | 2.415 |
| hdce6 | 3-methylcyclopentadiene | -42.34 | -43.44 | 2.415 |
| | Unlinked Ligands | — | — | 2.413 |

Unlike the thorium and protactinium complexes, the isoindene link makes only two appearances in the top twelve complexes for uranium. Additionally, isoindene is the only repeating link in the top twelve uranium complexes. The relaxed relative binding energies are much closer to the frozen relative binding energy than the protactinium complexes, but not quite as close as thorium. The largest decrease is nearly 6 kcal/mol in the rmsd6 complex. The complexes with the rmsd3 and hdce4 links actually destabilize by around 3 kcal/mol when the complex is allowed to relax. We suspect that it is related to the configuration of the link selected by HostDesigner. An additional difference between the uranium complexes and the previous actinide complexes is that the links did not all cause the average uranium-oxygen bond length to shorten, i.e. the links rmsd4, rmsd5, hdce1, hdce2, and hdce3 all cause the average uranium-oxygen bond length to lengthen.

Figure 13 is a graphical representation of the relaxed relative binding energy of the complexes shown in Table 5. In this case, the complexes containing isoindene links (hdce2 and hdce3) optimize to the same orientation, since their relaxed relative binding energies are approximately equivalent. The complex with the butene link (rmsd1) is the most stable by a significant amount, almost 47 kcal/mol. The remaining complexes are much closer in relative binding energy, with a range of 23 kcal/mol.

Additionally, it is worth noting that for uranium, the rmsd ranking method did a better job of predicting the most stable structures than the hdce algorithm. In both ranking methods, the most stable complex is ranked as the top structure, i.e. rmsd1 and hdce1. This is not the case for thorium or protactinium complexes. The results for uranium represent a best case scenario for the use of HostDesigner, and this might be related to more accurate force field parameters for uranium than for thorium and protactinium. If the algorithm could always predict the most stable structure as the first complex, the necessary computational time would be significantly reduced, since only two structures would need to be optimized, rather than twelve.

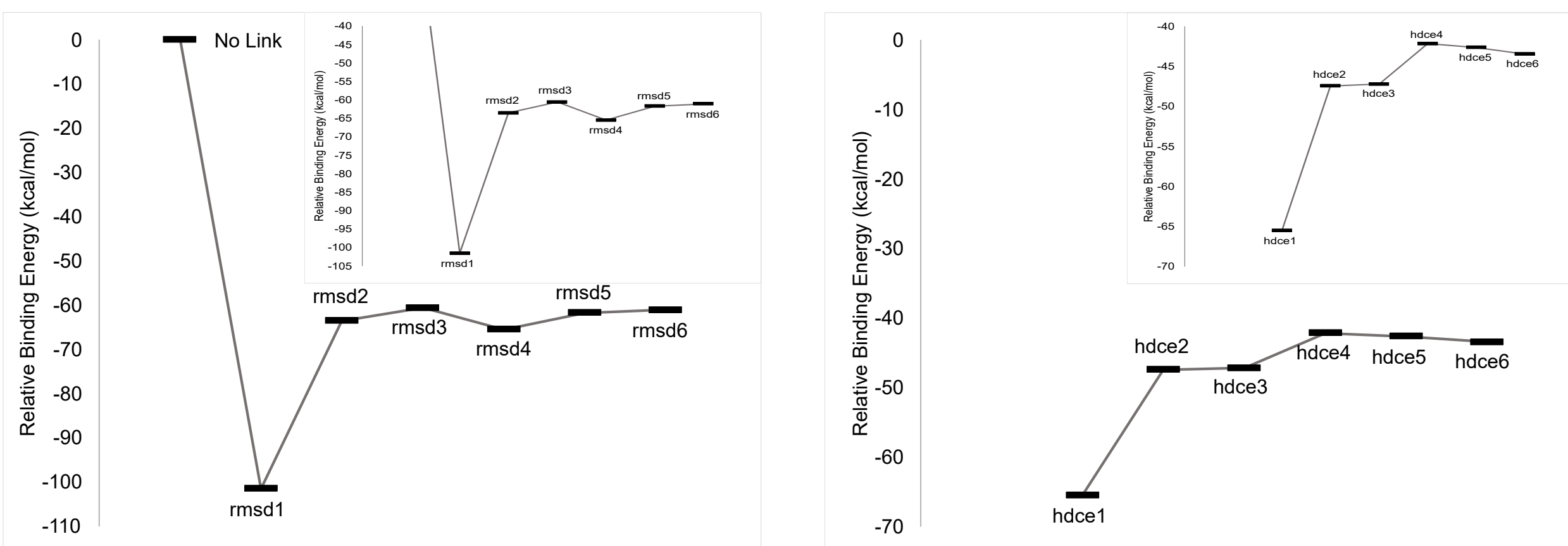

Figure 13: Relaxed binding energy of the top twelve uranium complexes produced by HostDesigner relative to the original, unlinked uranium-catecholamide complex.

In all cases, the addition of the link that connects two of the catecholamide ligands stabilizes the complex, i.e. all of the top twelve structures obtained from HostDesigner for inputs of thorium-catecholamide, protactinium-catecholamide, and uranium-catecholamide. The addition of covalent interactions to combine two existing ligands causes the resulting structures to be even more stable than the original complex.

# Conclusions

In this work, new chelating agents with the potential to facilitate the separation of radioactive waste are designed. We demonstrate that the secondary coordination sphere improves selectivity. An increase in interactions between ligands bound to a complex increases the overall stability of the complex as demonstrated through the investigation of relative binding energies of complexes containing actinide elements. In particular, the complexes with non-interacting ligands (actinide-catecholates) are stabilized through the addition of one amide group to each ligand. The resulting complexes (actinide-catecholamides) have non-covalent interactions between the ligands due to the amide groups. When the high-throughput screening algorithm, HostDesigner, is used to form covalent bonds (links) between ligands, the stablity of each complex increases. Molecular mechanics calculations are used by HostDesigner to obtain different configurations, choose and rank the links, while relativistic density functional theory (DFT-ZORA) calculations are used to further improve the geometry and to compute the binding energy of each complex.

DFT-ZORA is applied using the Amsterdam Density Functional (ADF) code. Gibbs free energy and average plutonium-oxygen bond lengths are determined for the plutonium-nitrate and plutonium-water complexes to validate this method. The results from calculations performed on these complexes are comparable to those found in literature. Therefore, this approach is capable of accurately predicting the optimal geometric structure of complexes containing actinide elements.

The actinide-catecholamide complexes optimized using ADF are used as input structures for HostDesigner. The top twelve structures created for each actinide are obtained from HostDesigner. In this work, geometry optimization calculations are only performed on complexes containing thorium, protactinium, and uranium. In all twelve complexes for each of these actinides, the addition of the link that connected two of the catecholamide ligands increases the overall stability of the complex. The addition of covalent interactions to combine two existing ligands causes the resulting structures to be more stable than the original complex with only non-covalent interactions among ligands. For thorium and protactinium, the most stable complex is generated by introducing an isoindene link, while for uranium the link that creates the most stability is butene.

Thus, we have presented a method that is feasible for designing actinide complexes that can be used for separating actinides. This method exploits the secondary sphere coordination effects, as well as the high-throughput screening of link and ligands, while using computational efficient relativistic DFT.

## Acknowledgement

This work was supported by startup funds from Florida State University (FSU). J.L.M-C. gratefully acknowledges the support from the Energy and Materials Initiative at FSU. The authors thank the High Performance Computer cluster at the Research Computing Center in FSU for providing computational resources and support.

# Supplementary Information

## Methods: Ligand Design Using HostDesigner

HostDesigner employs one of two algorithms to generate host molecules from user-defined fragments: LINKER or OVERLAY. If LINKER is called, a linking molecule from HostDesigner's library of "links", bridges the gap between two input fragments. If OVERLAY is called, a linking molecule from the library forms two bonds with one input fragment. Both OVERLAY and LINKER perform molecular mechanics calculations. The large, complex, and primarily multidentate nature of complexes containing actinide elements lends itself to the use of OVERLAY rather than LINKER. Therefore, only the OVERLAY algorithm is discussed in detail here. Figure 3 shows an example of how OVERLAY connects two existing ligands to form a more stable complex.

The first step a user takes in preparing HostDesigner is to create a "control" file. This file instructs HostDesigner to call one of the two algorithms, and defines specific variations of the program that the user may wish to apply. Below is an example of a control file.

Code 1: Control File for Host Designer

```
        OVER hosta=plutonium notype drivea
```

Code 1 calls the OVERLAY algorithm using the command "OVER". The input geometry is contained in a file called plutonium, defined using the command "hosta=plutonium". The "notype" command means that no atom type is specified in the input file (this is leftover from previous versions of HostDesigner, and has no impact on the ligand design). The command "drivea" calls geometry drives. Geometry drives allow for flexibility of the input structure, so that the orientation of the host may be changed to stabilize the resulting structures. For more information about geometry drives, see the HostDesigner User's Manual.[4] After the control file has been created, the user creates an input geometry file.

The structure of the input molecule is user-defined, and must be written to a file in a specific structure before executing HostDesigner. In the file, the user must specify coordinates (in a Cartesian system) for all atoms, their connectivity, and attachment points. Atom connectivity refers to the identification of atoms that have bonds to other atoms. Attachment points define the hydrogen atoms in the input fragment that will be replaced by carbon atoms from the linking fragment (link). Note that the OVERLAY algorithm requires the attachment points to be in pairs. Code 2, below, shows the "plutonium" file referenced in Code 1.

Code 2: HostDesigner Input Geometry File

```
Optimized plutonium-catecholate complex
77 1
O   1  -1.785  -0.223   1.569  77   2
C   2  -2.473   0.793   2.030   7   1   3
C   3  -3.455   0.702   3.011   2   4  49
C   4  -4.174   1.844   3.428   5   3  50
C   5  -3.923   3.074   2.861   6  51   4
C   6  -2.928   3.221   1.864  33   7   5
C   7  -2.176   2.089   1.454   8   6   2
O   8  -1.219   2.106   0.568   7  77
O   9  -1.063  -2.032  -0.848  10  77
C  10  -1.680  -1.969  -1.994  11  15   9
C  11  -2.269  -3.075  -2.661  12  37  10
C  12  -2.893  -2.878  -3.917  13  57  11
C  13  -2.942  -1.625  -4.488  56  12  14
C  14  -2.384  -0.511  -3.826  13  55  15
C  15  -1.763  -0.649  -2.586  14  10  16
O  16  -1.250   0.340  -1.897  15  77
O  17   0.933  -2.092   0.916  18  77
C  18   1.610  -2.062   2.029  17  23  19
C  19   2.208  -3.191   2.648  41  18  20
C  20   2.898  -3.028   3.874  19  63  21
C  21   3.002  -1.787   4.463  22  20  62
C  22   2.434  -0.650   3.848  23  61  21
C  23   1.750  -0.757   2.641  24  18  22
O  24   1.218   0.251   1.991  77  23
O  25   1.687  -0.399  -1.436  26  77
C  26   2.396   0.564  -1.968  27  31  25
C  27   3.362   0.378  -2.955  28  67  26
C  28   4.115   1.464  -3.448  68  29  27
C  29   3.916   2.737  -2.957  28  69  30
C  30   2.938   2.981  -1.964  29  31  45
C  31   2.151   1.904  -1.476  26  30  32
O  32   1.207   2.013  -0.583  31  77
C  33  -2.754   4.560   1.246  35  34   6
O  34  -3.458   5.551   1.527  33
N  35  -1.766   4.636   0.316  36  73  33
C  36  -1.618   5.789  -0.528  54  52  53  35
C  37  -2.311  -4.436  -2.068  11  38  39
O  38  -2.890  -5.408  -2.597  37
N  39  -1.675  -4.558  -0.873  37  75  40
C  40  -1.808  -5.737  -0.064  39  60  59  58
C  41   2.198  -4.541   2.028  43  42  19
O  42   2.774  -5.534   2.518  41
N  43   1.523  -4.623   0.852  44  74  41
C  44   1.614  -5.783   0.009  66  64  65  43
C  45   2.814   4.363  -1.434  30  46  47
O  46   3.552   5.308  -1.779  45
N  47   1.830   4.536  -0.512  45  76  48
C  48   1.714   5.752   0.242  47  71  72  70
H  49  -3.654  -0.273   3.443   3
H  50  -4.935   1.744   4.197   4
H  51  -4.475   3.959   3.149   5
H  52  -2.533   5.997  -1.097  36
H  53  -1.391   6.686   0.057  36
H  54  -0.798   5.602  -1.219  36
H  55  -2.429   0.481  -4.264  14
H  56  -3.419  -1.486  -5.455  13
H  57  -3.331  -3.743  -4.399  12
H  58  -1.175  -5.620   0.814  40
H  59  -2.845  -5.901   0.258  40
```

```
62  H  60   -1.496  -6.629  -0.614   40
63  H  61    2.523   0.333   4.298   22
64  H  62    3.532  -1.676   5.405   21
65  H  63    3.344  -3.909   4.317   20
66  H  64    2.641  -5.961  -0.335   44
67  H  65    1.290  -6.683   0.539   44
68  H  66    0.970  -5.627  -0.855   44
69  H  67    3.520  -0.628  -3.327   27
70  H  68    4.863   1.290  -4.217   28
71  H  69    4.496   3.582  -3.305   29
72  H  70    0.882   5.644   0.936   48
73  H  71    1.524   6.610  -0.409   48
74  H  72    2.630   5.971   0.806   48
75  H  73   -1.307   3.736   0.095   35
76  H  74    1.154  -3.725   0.499   43
77  H  75   -1.301  -3.674  -0.490   39
78  H  76    1.338   3.670  -0.236   47
79  Pu 77   -0.041   0.001   0.065   16   25   9   32   1   24   8   17
80  8
81   53   C   0
82   70   C   0
83   53   C   0
84   71   C   0
85   54   C   0
86   70   C   0
87   54   C   0
88   71   C   0
89  D  2
90   D  I  -20.  20.  10.  48  47
91   D  I  -20.  20.  10.  36  35
92  END
```

The first line of the input file is an arbitrary, user-defined title. Line 2 indicates the number of atoms that are in the structure, 77, and the number of atoms in the guest (metal), 1. Lines 3 - 79 define each atom in the complex. The first column is the chemical element symbol for each atom, the second column is each atom's assigned serial number, the third, fourth, and fifth are the (x,y,z) coordinates of each atom, and the rest of the columns are the atom connectivity given by their respective serial numbers. After all the atoms in the structure are defined, the attachment points are selected. Line 80 specifies the number of potential attachment points, 81 - 88 indicate the hydrogen atoms (in pairs) that will be replaced with carbon atoms and connected by HostDesigner using a link. Finally, lines 89 - 91 specify the geometry drives that were mentioned above.

Figure 2 shows an overview of the steps that HostDesigner's OVERLAY algorithm takes to design potential chelating agents for a given input structure: in other words, how Figure 3a becomes Figure 3b. OVERLAY checks HostDesigner's library for a suitable link based on the input geometry. After examining the link and its connection to the host molecule, OVERLAY decides if the link satisfies the specified parameters. If there is no match, OVERLAY returns to the library and selects another link. If there is a match, the coordinates of the link are added to those of the host molecule, and are written to the appropriate output file. This process is repeated $n$ times, for every link in the library, currently 8,266 links. Finally, once both the input and control files are complete, HostDesigner may be run using the command "hd3.0".

After HostDesigner runs, it provides two output files containing integrated host-guest complexes ranked in two different ways. Broadly speaking, in the first output file the structures are ranked based on geometric parameters (rmsd), and in the second output file based on estimated conformational energy parameters (hdce). The output files are provided as a list of .xyz files, therefore it is straightforward to copy and paste the coordinates into a visualization tool. A more in-depth explanation of ranking methods, and a step-by-step description of the OVERLAY algorithm can be found in the HostDesigner User's Manual.[4]

Several points about HostDesigner must be emphasized. OVERLAY and LINKER algorithms are intended for the preliminary design of new host molecules. More detailed analysis of output structures must follow the use of HostDesigner. Also, the order of the atoms in the input file is not arbitrary. The atoms must be listed in serial order, with the non-hydrogen atoms first, the hydrogens next, and the guest last. To sort atoms, it is convenient to use the open-source molecular builder and visualization tool, Avogadro, which allows for sorting of atoms using the atom properties tool.[3]

Once the atoms are sorted properly in Avogadro, the coordinates are exported as a .bgf file. Because the .bgf format includes information that is unnecessary for HostDesigner, the file must be rearranged and edited to match the format discussed above. Finally, input files are currently subject to the following limitations:

1. The atom lost from the input fragment to form a bond to the link must be a hydrogen

2. The connecting atom from the link must be carbon

3. The hydrogen can only be lost from:

    - Alkane, alkene, and arene carbons

    - Aliphatic alcohol and phenol oxygens

- Amine and amide nitrogens

The limitations of the HostDesigner algorithm require the user to be selective when choosing an input structure. For the purposes of this work, it was straightforward to begin with one of the complexes that was provided in the HostDesigner download package; uranium bound to four catecholamide ligands. In this way, all the constraints were known to be satisfied.